\documentclass[artice,aps,pra,showpacs,twocolumn,superscriptaddress]{revtex4}
\usepackage{graphicx,color}
\usepackage{amsmath,amsthm,amsfonts,amssymb,bm}
\usepackage{times}
\usepackage{epsf}
\usepackage[colorlinks={true}]{hyperref}
\usepackage[T1]{fontenc}
\usepackage[utf8]{inputenc}
\hypersetup{citecolor={blue}, filecolor={blue}, linkcolor={blue}, urlcolor={blue}}

\newcommand{\commentold}[1]{}
\DeclareMathSymbol{:}{\mathpunct}{operators}{"3A}
\usepackage[utf8]{inputenc}
\usepackage[english]{babel}
\usepackage{amsthm}
\theoremstyle{definition}

\begin{document}

\title{Protecting the entropic uncertainty lower bound in Markovian and
non-Markovian environment via additional
qubits}
\author{S. Haseli}
\email{soroush.haseli@uut.ac.ir}
\affiliation{Faculty of Physics, Urmia University of Technology, Urmia, Iran
}
\author{F. Ahmadi}
\affiliation{Department of Engineering Science and Physics, Buein Zahra Technical University
}


\date{\today}

\begin{abstract}
The uncertainty principle is an important principle in quantum theory. Based on this principle, it is impossible to predict the measurement outcomes of two incompatible observables, simultaneously. Uncertainty principle basically is expressed in terms of the standard deviation of the measured observables. In quantum information theory, it is shown that  the uncertainty principle can be expressed by  Shannon's entropy. The entopic uncertainty lower bound can be altered by considering a particle as the quantum memory  which is correlated with the measured particle.  We assume that the quantum memory is an open system. We also select the quantum memory from $N$ qubit which interact with common reservoir. In this work we investigate the effects of the number of additional qubits in reservoir on entropic uncertainty lower bound. We conclude that the entropic uncertainty lower bound can be protected from decoherence by increasing the number of additional qubit in reservoir.
\end{abstract}

\maketitle
\section{Introduction}\label{Sec1}
Uncertainty principle is  one of the most fundamental and important issues in quantum theory which is determined the distinction between quantum theory and classical theory. It
sets a nontrivial limit on our ability to predict the 
outcomes of two incompatible measurements on a quantum system.  The first and well-known uncertainty relation was proposed by Heisenberg\cite{Heisenberg}. It is associated with momentum  and position, measured for a particle. Based on the standard deviation of
measuring outcomes of the incompatible observables position  $\hat{x}$ and momentum $\hat{p}$, the Heisenberg uncertainty relation is formulated by Kennard as \cite{Kennard}
\begin{equation}
\Delta \hat{p} \Delta \hat{x} \geq \frac{\hslash}{2}.
\end{equation}
Robertson \cite{Robertson} and Schrodinger  \cite{Schrodinger},  improved the above uncertainty relation for arbitrary pairs of incompatible observables $\hat{Q}$ and $\hat{R}$ to a universal form as
\begin{equation}
\Delta \hat{Q} \Delta \hat{R} \geq \frac{1}{2} \vert \langle \psi \vert \left[ \hat{Q},\hat{R} \right] \vert \psi \rangle \vert,
\end{equation}
where $\Delta \hat{O} = \sqrt{ \langle \psi \vert \hat{O}^{2} \vert \psi \rangle - \langle \psi \vert \hat{O} \vert \psi \rangle^{2}}, \quad (\hat{O} \in \lbrace \hat{Q}, \hat{R} \rbrace) $ is  the standard deviations and $\left[\hat{Q},\hat{R} \right] = \hat{Q}~\hat{R}-\hat{R}~\hat{Q}$. The lower bound  of Robertson's uncertainty relation depends on the state of the system. It leads to a trivial result when the system is prepared in the eigenstates of anyone of the two observables. With the advent of quantum information theory, it was observed that the relation of uncertainty can be formulated in terms of Shannon entropy. The first entropic uncertainty relation EUR was suggested by Kraus \cite{Kraus}, and then was proved by Maassen and Uffink \cite{Maassen}
\begin{equation}\label{entropic1}
H(Q)+H(R) \geq \log_{2}\frac{1}{c},
\end{equation} 
where $H(\hat{O})=-\sum_{o} p_{o} \log_{2} p_{o}$ is the Shannon entropy of the measured observable $\hat{O} \in \lbrace Q, R\rbrace$, $p_o$  is the probability of
the outcome $o$, and $c=\max_{\lbrace i,j\rbrace} \vert \langle q_i \vert r_j \rangle \vert^{2}$ quantifies the complementarity between the observables. If the state of the  measured particle is $\rho$, then the EUR can be formulated in general form as 
\begin{equation}\label{entropic2}
H(Q)+H(R) \geq \log_2 \frac{1}{c}+S(\rho)
\end{equation} 
where $S(\rho)=-tr(\rho \log_2 \rho)$ is the von Neumann entropy of the general state $\rho$. When the state $\rho$ is pure, the above EUR reduces to Eq. \ref{entropic1}. In general, the EUR can be interpreted by an interesting game between two observers, Alice and Bob. At firts, Bob prepares
a particle in an arbitrary quantum state $\rho$ and sends it to Alice. Then, they reach an agreement on  measuring two observables $\hat{Q}$ and $\hat{R}$ by Alice on the prepared particle. Alice does her measurement on the quantum state of the prepared particle $\rho$ and declares
her choice of the measurement to Bob. Bob tracks to minimize his uncertainty about the outcome of
Alice  measurement . The  minimum of the uncertainty of Bob about Alice's measurement is bounded by  Eq. \ref{entropic2}. Let us consider the game with two particles. Bob prepares a correlated bipartite state
$\rho_{AB}$ and sends one of them to Alice and the other one is kept
as the quantum memory.  In this game, the minimum of the uncertainty of Bob about Alice's measurement is bounded by the Quantum-memory-assisted EUR as \cite{Berta}
\begin{equation}\label{berta}
S(Q \vert B)+ S(R \vert B) \geq \log_2 \frac{1}{c} + S(A \vert B),
\end{equation}
where $S(O \vert B)=S(\rho^{OB})-S(\rho^{B})$ denotes  the conditional von Neumann entropies of the post measurement states
\begin{equation}
\rho^{OB}=\sum_{i} (\vert o_i \rangle \langle o_i \vert \otimes \mathbf{I})\rho^{AB} (\vert o_i \rangle \langle o_i \vert\otimes \mathbf{I}),
\end{equation}
where $\lbrace \vert o_i \rbrace$'s are the eigenstates of the observable $O$, and $\mathbf{I}$ is the identity operator. In comparison with Compared with Eqs. \ref{entropic1} and \ref{entropic2}, one can see that the uncertainty lower bound in Eq. \ref{berta} reduces for the negative conditional entropy $S(A \vert B)$. So, Bob can guess Alice's measurement outcomes
with better accuracy. Quantum-memory-assisted EUR, has a wide range of applications including entanglement detection \cite{Huang,Prevedel,Chuan-Feng} and quantum cryptography \cite{Tomamichel,Ng}.

 Much efforts have been made to find the tighter uncertainty bounds for Quantum-memory-assisted EUR \cite{Pati,Pramanik,Coles,Liu,Zhang,Pramanik1,Adabi,Adabi1,Dolatkhah,Jin-Long}. In Ref.\cite{Adabi} Adabi et al. proposed a tighter  lower bound for Quantum-memory-assisted EUR. They showed that the
uncertainty bound of Eq. \ref{berta} can be tightened as
\begin{equation}\label{concen}
S(Q\vert B)+S(R\vert B) \geq \log_2 \frac{1}{c} + S(A \vert B)+ \max \lbrace 0, \delta\rbrace,
\end{equation}
where 
\begin{equation}
\delta = I(A;B)-(I(Q;B)+I(R;B)),
\end{equation}
and
\begin{equation}\label{holevo}
I(\hat{O};B)=S(\rho^{b})-\sum_{i}p_{i}S(\rho_{i}^{B})
\end{equation}
is the Holevo quantity. It shows the Bob's accessible information about Alice's measurement $\hat{O}$. When Alice measures observable $\hat{O}$, the $i$-th outcome with probability $p_i= tr_{AB}(\Pi_i^{A} \rho^{AB} \Pi_i^{A}) $ is obtained and Bob state  is left in the corresponding state $\rho_i^{B}=\frac{tr_{A}(\Pi_i^{A} \rho^{AB} \Pi_i^{A})}{p_i}$. Adabi's uncertainty bound given on the right-hand side (RHS) of Eq. \ref{concen} is tighter than other bound which have introduced by others. 

In a real sense, quantum systems interact with their surrounding subjected
to information loss in the form of dissipation and decoherence. 
In this work we consider the case in which $N$ qubits interacts with common environment.  We select the quantum memory from these $N$ qubits. We investigate how these additional qubits effect on entropic uncertainty lower bound EULB. The work is organized as follow. In Sec. \ref{Sec2}, We will review the dynamical model which is used in this work.  We will examine some
examples in Sec. \ref{Sec3}. The manuscript will close with a conclusion
in Sec. \ref{Sec4}.
\section{Dynamical model}\label{Sec2}
Let us consider $N$ single-qubit which are located in a common  dissipative reservoir. We suppose that each qubit is independently coupled to common zero temperature thermal reservoir that is consist of harmonic oscillators.  The Hamiltonian of the total system ($N$ single qubit + reservoir) is given by \cite{Behzadi1,Behzadi2}
\begin{eqnarray}
H&=&\hat{H}_0 + \hat{ H}_I \nonumber \\
&=& \omega_0 \sum_{i=1}^{N} \sigma_i^{+} \sigma_{i}^- + \sum_k \omega_{k}b_{k}^\dag b_{k} \nonumber \\
&+& \sum_{i=1}^{k}\sum_{k}(g_k^{\star}b_k^{\dag} \sigma_{i}^{-}+g_k b_k \sigma_{i}^{+}),
\end{eqnarray}
where $g_k$'s are the coupling strength between the $i^{th}$ qubit with transition frequency $\omega_0$ and $k^{th}$ field mode with frequency $\omega_k$, $b_k^{\dag}$ and $b_k$ are creation and annihilation operators of the $k^{th}$ field mode, respectively. $\sigma_i^{+}$ and $\sigma_i^{-}$ are the $i^{th}$ qubit  raising and lowering operators , respectively. We consider the case in which there exist one excitation in the total system and reservoir in the vacuum state $\vert 0 \rangle_E$, initially. We also suppose that the initial state of the whole system is given by
\begin{equation}
\vert \psi(0) \rangle =C_{0}(0) \vert 0 \rangle_s \otimes \vert 0 \rangle_E+ \sum_{i=1}^{N} C_i(0)\vert i \rangle_s \otimes \vert 0 \rangle_E,
\end{equation}
where $\vert 0 \rangle_s$ means that all qubits are in ground state $\vert 0 \rangle$, and  $\vert i \rangle_s$ shows that $i^{th}$ qubit in the excited state $\vert 1 \rangle$ and the others are in ground state $\vert 0 \rangle$.  The dynamics of the whole system can be written as  
 \begin{eqnarray}
\vert \psi(t) \rangle &=& (C_0(t) \vert 0 \rangle_s + \sum_{i=1}^{N} C_i(t) \vert i \rangle_s) \otimes \vert 0 \rangle_E \nonumber \\
&+& \sum_{j=1}^{N} C_{j}(t) \vert 0 \rangle_s \otimes \vert 1_j \rangle_E,
 \end{eqnarray}
 where $\vert 1_j \rangle_E$ is the state of the reservoir with single excitation in the $j^{th}$ field mode. The Schrodinger equation in the interaction picture has the form 
 \begin{equation}\label{sch}
 i \frac{d}{d t}\vert \psi(t) \rangle = \hat{H}_I(t) \vert \psi(t) \rangle,
 \end{equation}
 where $\hat{H}_I(t)=e^{i \hat{H}_0 t}\hat{H}_I e^{-i \hat{H}_0 t}$. By solving Eq. (\ref{sch}) and following the method which has outlined in Ref. \cite{Behzadi2}, the dynamis of the $i^{th}$ qubit can be obtained as 
 \begin{equation}
 \rho_i(t)=\left(
\begin{array}{cc}
 \rho_{00}^{i} \vert C_i(t) \vert ^{2}  & \rho_{01}^{i} C_{i}(t) \\
 \rho_{10}^{i} C_{i}^{\star}(t) &  1-\rho_{00}^{i} \vert C_i(t) \vert ^{2}  \\
\end{array}
\right).
\end{equation}  
 From Eq. (\ref{sch}), we have $\dot{C}_0(t)=0$ and so $C_0(t)=C_0(0)=C_0$. The other probability coefficients satisfies the following integro-differential equations
 \begin{equation}\label{cof}
 \frac{d}{dt}C_{i}(t)=-\int_0^{\infty} \int_{-\infty}^{+ \infty} J(\omega)e^{i(\omega_0-\omega_k)(t-\tau)}\sum_{j=1}^{N}C_{j}(\tau) d \omega d \tau,
 \end{equation}
 where $J(\omega)$ is the spectral density of the reservoir.  Let us consider a Lorantzian spectral density 
 \begin{equation}
 J(\omega)=\frac{1}{2 \pi}\frac{\gamma_0 \lambda^2}{(\omega_0-\omega)^{2} + \lambda^{2}},
 \end{equation}
where the spectral width of the coupling $\lambda$ is related  to the correlation time of the environment $\tau_E$ via $\tau_E \simeq 1/ \lambda$. The parameter $\gamma_0$ is connected to the relaxation time $\tau_E$, over which the state of the system changes, by $\tau_E \simeq 1/\gamma_0$. If $\gamma_0 /\lambda \leq 1/2$ we have the weak system-reservoir coupling regime and the dynamic is Markovian, while for  $\gamma_0 /\lambda > 1/2$, we have the strong coupling regime and the dynamic is on-Markovian. Using Laplace transformation and its inverse, the probability coefficient $C_i(t)$ can be obtained as 
\begin{equation}
C_{i}(t)=\frac{N-1}{N}+\frac{e^{-\lambda t /2}}{N}[\cosh(\frac{D t}{2})+\frac{\lambda}{D}\sinh(\frac{D t}{2})],
\end{equation}
where $D=\sqrt{\lambda^{2}-2 N \gamma_0 \lambda}$.
\section{EULB and additional qubits}\label{Sec3}
In this section we use the above mentioned dynamical model to reduce the EULB in the presence of decoherence. Let us suppose that Bob prepare the initial correlated state $\rho^{AB_i}$ such that he chooses $i^{th}$ single-qubit from $N$ qubits in common reservoir as the part $B$. Then, he sends particle $A$ to Alice and keeps $B$ as the quantum memory.   Then, they reach an agreement on  measuring  observables  by Alice on her particle. Alice does her measurement on the quantum state of the prepared particle and declares
her choice of the measurement to Bob. Bob tracks to minimize his uncertainty about the outcome of
Alice  measurement . 

\begin{figure}[!]  
\centerline{\includegraphics[scale=0.5]{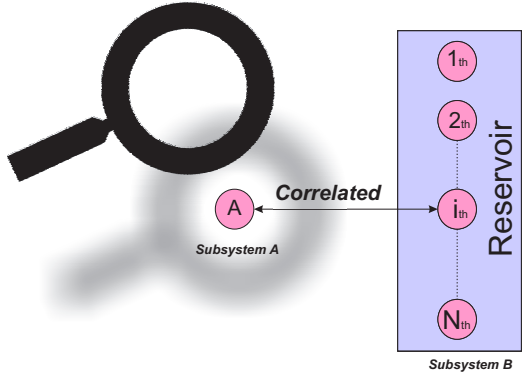}}
\caption{Schematic representation of the model where quantum memory interacts with environment. While Alice performs measurement on her particle. }\label{Fig1}
\end{figure}

In this model the quantum memory $B_i$ interacts with reservoir as an open quantum system (The model is sketched in Fig. (\ref{Fig1})). It is expected that, as a result of the interaction between quantum memory and  reservoir, the correlation between quantum memory $B$ and measured particle $A$ decreases. As mentioned before, the presence of correlation between Alice and Bob reduces the uncertainty of Bob about the outcomes of Alice's measurement. So, it is logical to expect that the EULB increases due to interaction between quantum memory and reservoir.  
\subsection{Examples}
\subparagraph{Maximally entangled state:} Let us consider the case in which Bob prepares a maximally entangled pure state $\vert \phi \rangle_{AB_i}=1/\sqrt{2}(\vert 0 0 \rangle_{AB_i} + \vert 11 \rangle_{AB_i})$.  If the quantum memory $B_i$ interacts  with environment, the dynamic of the bipartite quantum state $\rho_{AB_i}=\vert \phi \rangle_{AB_i} \langle \phi \vert$ can be obtained as 
 \begin{equation}
 \rho_i(t)=\frac{1}{2}\left(
\begin{array}{cccc}
 \vert C_i(t) \vert^{2}  & 0 & 0 & C_i(t) \\
 0 &  1-\vert C_i(t) \vert ^{2} &0&0  \\
 0 & 0 & 0 & 0 \\
 C_i^{\star}(t) & 0 & 0 & 1 \\
\end{array}
\right).
\end{equation}  
Alice and Bob reach an agreement on  measuring two observables $\hat{\sigma_x}$ and $\hat{\sigma_z}$. The von Neumann entropy of the post measurement states are given by
\begin{eqnarray}
S(\rho_{\hat{\sigma_{x}}B_i})&=&-\frac{1-\eta}{2}\log_2 (\frac{1-\eta}{4})-\frac{1+\eta}{2}\log_2 (\frac{1+\eta}{4})  \\
S(\rho_{\hat{\sigma_{z}}B_i})&=&\frac{1}{2}-\frac{\vert C_i(t) \vert^{2}}{2}\log_2 \frac{\vert C_i(t) \vert^{2}}{2} \nonumber \\
&-&\frac{1-\vert C_i(t) \vert^{2}}{2}\log_2 \frac{1-\vert C_i(t) \vert^{2}}{2},
\end{eqnarray}
 where $\eta = \sqrt{1-\vert C_i(t) \vert^{2}+\vert C_i(t) \vert^{4}}$. So, the left-hand side (LHS) of Eq.(\ref{concen}) is obtained as 
 \begin{eqnarray}
 U_L&=&\frac{1}{2}-\frac{1-\eta}{2}\log_2 (\frac{1-\eta}{4})-\frac{1+\eta}{2}\log_2 (\frac{1+\eta}{4}) \nonumber \\
 &-&\frac{\vert C_i(t) \vert^{2}}{2}\log_2 \frac{\vert C_i(t) \vert^{2}}{2}-\frac{1-\vert C_i(t) \vert^{2}}{2}\log_2 \frac{1-\vert C_i(t) \vert^{2}}{2} \nonumber \\
 &-&S_{bin}(\frac{\vert C_i(t) \vert^{2}}{2}),
 \end{eqnarray}
where $S_{bin}(x)=-x \log_2 x -(1-x)\log_2(1-x)$.  The right-hand side (RHS) of Eq.(\ref{concen}) is given by
 \begin{equation}
 U_R=1+S_{bin}(\frac{1-\vert C_i(t) \vert^{2}}{2})-S_{bin}(\frac{\vert C_i(t) \vert^{2}}{2})+\max\lbrace0,\delta\rbrace,
 \end{equation}
 where 
 \begin{eqnarray}
 \delta&=&-\frac{1}{2}-\frac{1-\eta}{2}\log_2 (\frac{1-\eta}{4})-\frac{1+\eta}{2}\log_2 (\frac{1+\eta}{4}) \nonumber \\ 
 &-&\frac{\vert C_i(t) \vert^{2}}{2}\log_2 \frac{\vert C_i(t) \vert^{2}}{2}-\frac{1-\vert C_i(t) \vert^{2}}{2}\log_2 \frac{1-\vert C_i(t) \vert^{2}}{2} \nonumber \\
 &-&S_{bin}(\frac{1-\vert C_i(t) \vert^{2}}{2})-S_{bin}(\frac{\vert C_i(t) \vert^{2}}{2}).
 \end{eqnarray}
 \begin{figure}[!]  
\centerline{\includegraphics[scale=0.5]{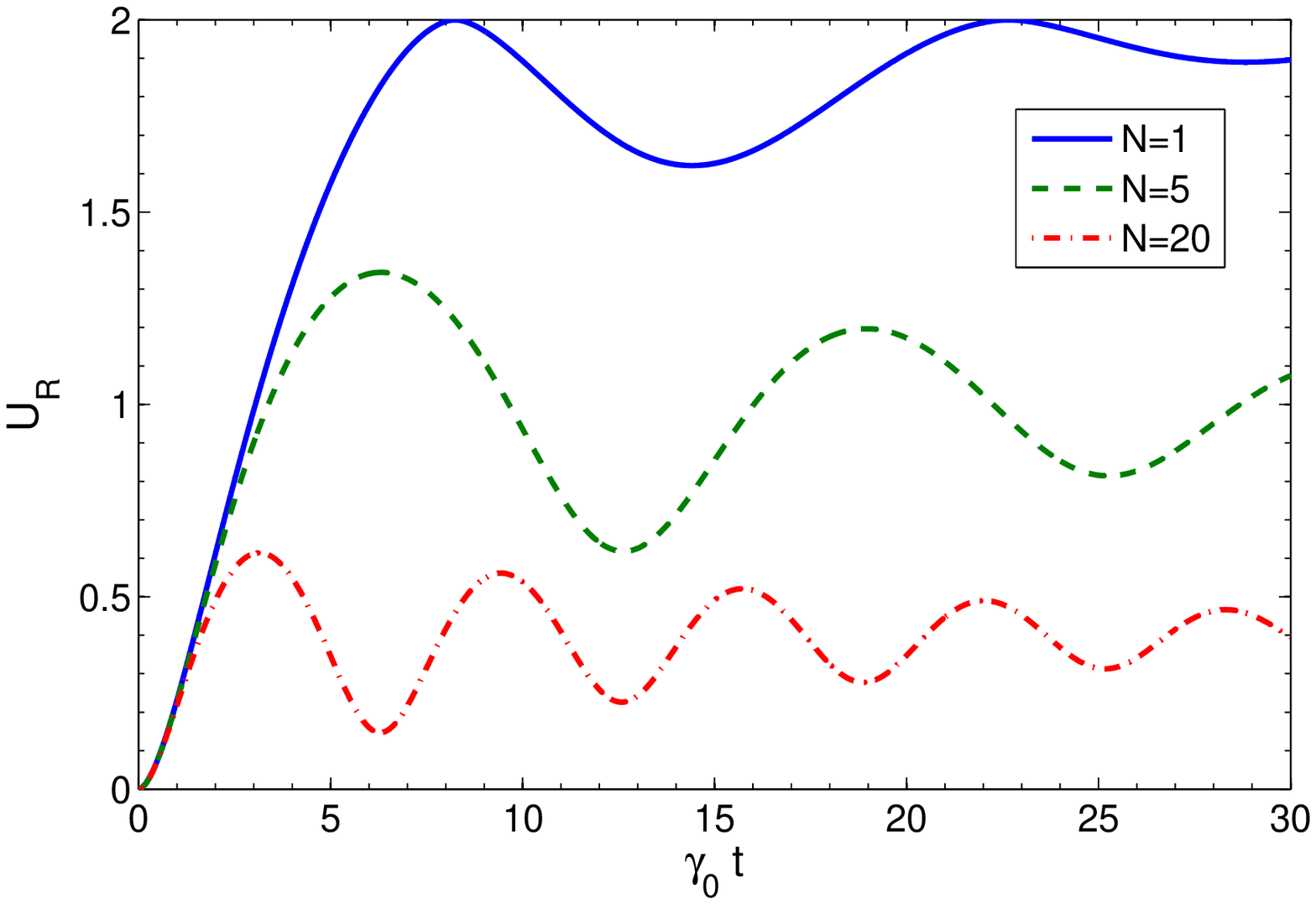}}
\caption{Lower bounds of the entropic uncertainty relation of the two complementary observables $\hat{\sigma}_x$ and $\hat{\sigma}_z$ as a function of $\gamma_0 t$, when Bob prepare a maximally entangled state $\vert \phi \rangle = 1/\sqrt{2}(\vert 00 \rangle + \vert 11 \rangle)$ in non-Markovian regime $\lambda=0.1 \gamma_0$. }\label{Fig2}
\end{figure}
\begin{figure}[!]  
\centerline{\includegraphics[scale=0.5]{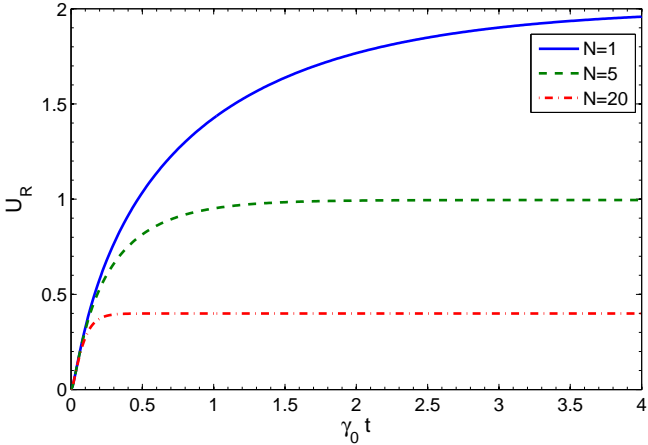}}
\caption{Lower bounds of the entropic uncertainty relation of the two complementary observables $\hat{\sigma}_x$ and $\hat{\sigma}_z$ as a function of $\gamma_0 t$, when Bob prepare a maximally entangled state $\vert \phi \rangle = 1/\sqrt{2}(\vert 00 \rangle + \vert 11 \rangle)$ in Markovian regime $\lambda=40 \gamma_0$.}\label{Fig3}
\end{figure}
In Fig. \ref{Fig2}, the EULB, $U_R$ is plotted as a function of time in non-Markovian regime $\lambda=0.1 \gamma_0$ for various number of additional qubit $N$. As can be seen from Fig. \ref{Fig2}, due to the interaction between quantum memory and reservoir, EULB is increased through time . One can see, the lower bound decreases by increasing the number of additional qubit . In Fig. \ref{Fig3},  $U_R$ is plotted as a function of time in Markovian regime $\lambda = 40 \gamma_0$ for various number of additional qubits $N$. As can be seen from Fig. \ref{Fig4}, $U_R$ is increased over time, while it is decreased by increasing the number of additional qubits. 
 \subparagraph{Bell diagonal state:} As the second example, let us consider the set of two-qubit states
with the maximally mixed marginal states. This state can be
written as
\begin{equation}\label{bell}
\rho_{AB_i}=\frac{1}{4}(\mathbf{I} \otimes \mathbf{I}+ \sum_{i=1}^{3}r_i \sigma_i \otimes \sigma_i),
\end{equation}
where $\sigma_i$'s are Pauli matrices and $\vec{r}=(r_1,r_2,r_3)$ belongs to a tetrahedron defined by the set of  $(-1,-1,-1)$ , $(-1,1,1)$, $(1,-1,1)$ and $(1,1,-1)$. Bob prepares the state with  $r_1=1-2p$, $r_2=r_3=-p$,  such that the state in Eq. \ref{bell} becomes
\begin{equation}
\rho^{AB_i}=p \vert \psi^- \rangle\langle \psi^- \vert + \frac{1-p}{2}(\vert \psi^+ \rangle\langle \psi^+ \vert + \vert \phi^+ \rangle\langle \phi^+ \vert),
\end{equation} 
where $\vert \phi^{\pm} \rangle = \frac{1}{\sqrt{2}}[\vert 00 \rangle \pm \vert 11 \rangle]$ and $\vert \psi^{\pm} \rangle = \frac{1}{\sqrt{2}}[\vert 01 \rangle \pm \vert 10 \rangle]$ are the Bell diagonal states. In the following the Bell-diagonal state with $p=1/2$ is considered. 
The dynamics of density matrix when quantum memory $B_i$ interacts with reservoir is given by
\begin{equation}
\rho_{AB_i}=\left(
\begin{array}{cccc}
 \rho_{11}^{t} & 0 & 0 & \rho_{14}^{t} \\
 0 & \rho_{22}^{t} & \rho_{23}^{t} & 0 \\
 0 & \rho_{32}^{t} & \rho_{33}^{t} & 0 \\
 \rho_{41}^{t} & 0 & 0 & \rho_{44}^{t} \\
\end{array}
\right)
\end{equation}
where 
\begin{eqnarray}
\rho_{11}^{t}&=&\frac{1+p}{4} \vert C_i(t) \vert^{2}, \quad \rho_{14}^{t}=\frac{1-p}{2} \vert C_i(t) \vert \nonumber \\
\rho_{22}^{t}&=&\frac{1-p}{4}+\frac{1+p}{4} \left(1- \vert C_i(t) \vert ^{2} \right), \quad \rho_{23}^{t}=\frac{1-3p}{2}  C_i(t) \nonumber \\
\rho_{33}^{t}&=&\frac{1}{4} (1-p) \vert C_i(t) \vert^{2}, \quad \rho_{44}^{t}=\frac{1+p}{4}+\frac{1}{4} (1-p) \left(1-\vert C_i(t) \vert^{2}\right). \nonumber 
\end{eqnarray}
 \begin{figure}[!]  
\centerline{\includegraphics[scale=0.5]{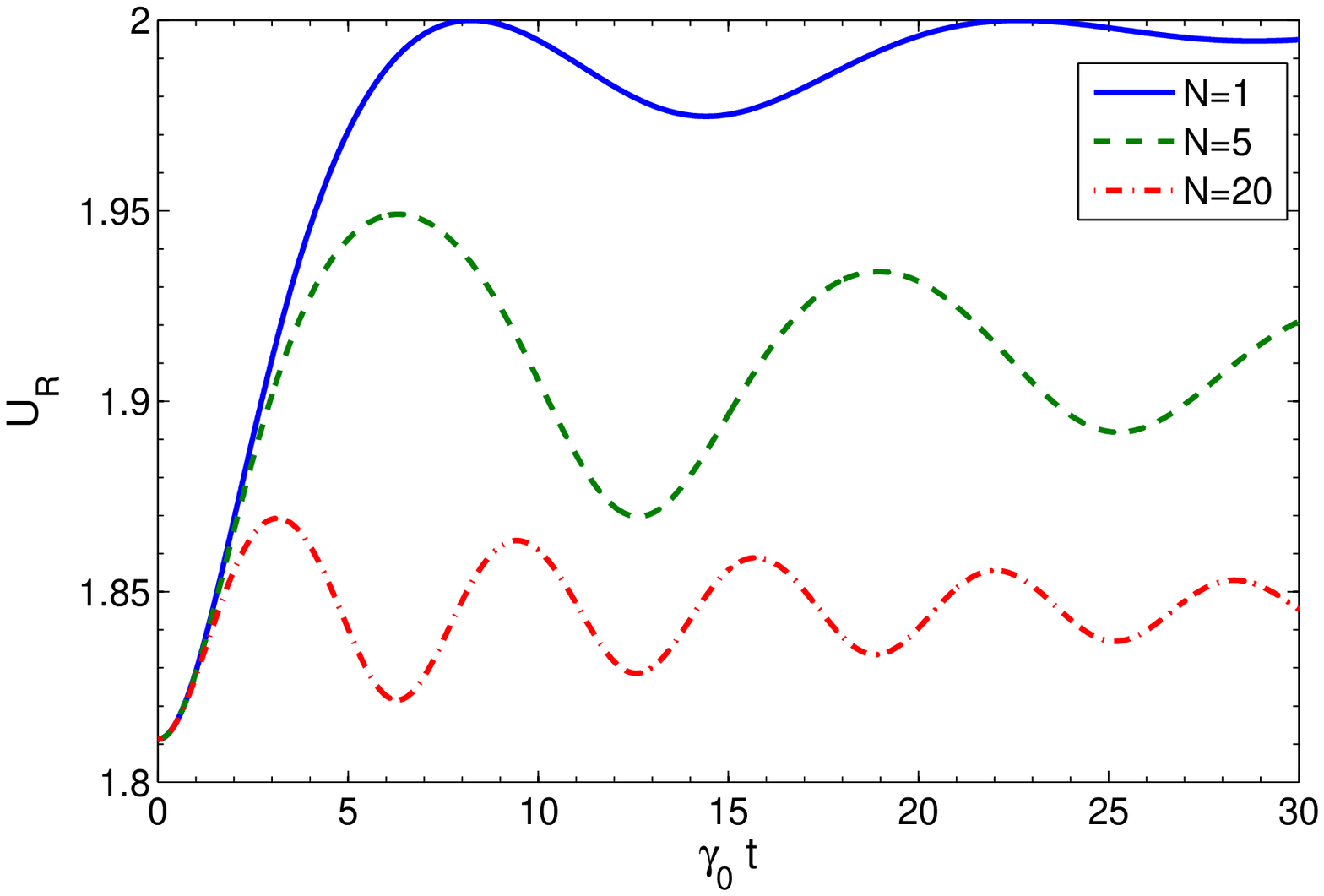}}
\caption{Lower bounds of the entropic uncertainty relation of the two complementary observables $\hat{\sigma}_x$ and $\hat{\sigma}_z$ as a function of $\gamma_0 t$, when Bob prepare a Bell diagonal state $\rho^{AB_i}=p \vert \psi^- \rangle\langle \psi^- \vert + \frac{1-p}{2}(\vert \psi^+ \rangle\langle \psi^+ \vert + \vert \phi^+ \rangle\langle \phi^+ \vert)$ with $p=1/2$ in non-Markovian regime $\lambda=0.1 \gamma_0$. }\label{Fig4}
\end{figure}
\begin{figure}[!]  
\centerline{\includegraphics[scale=0.5]{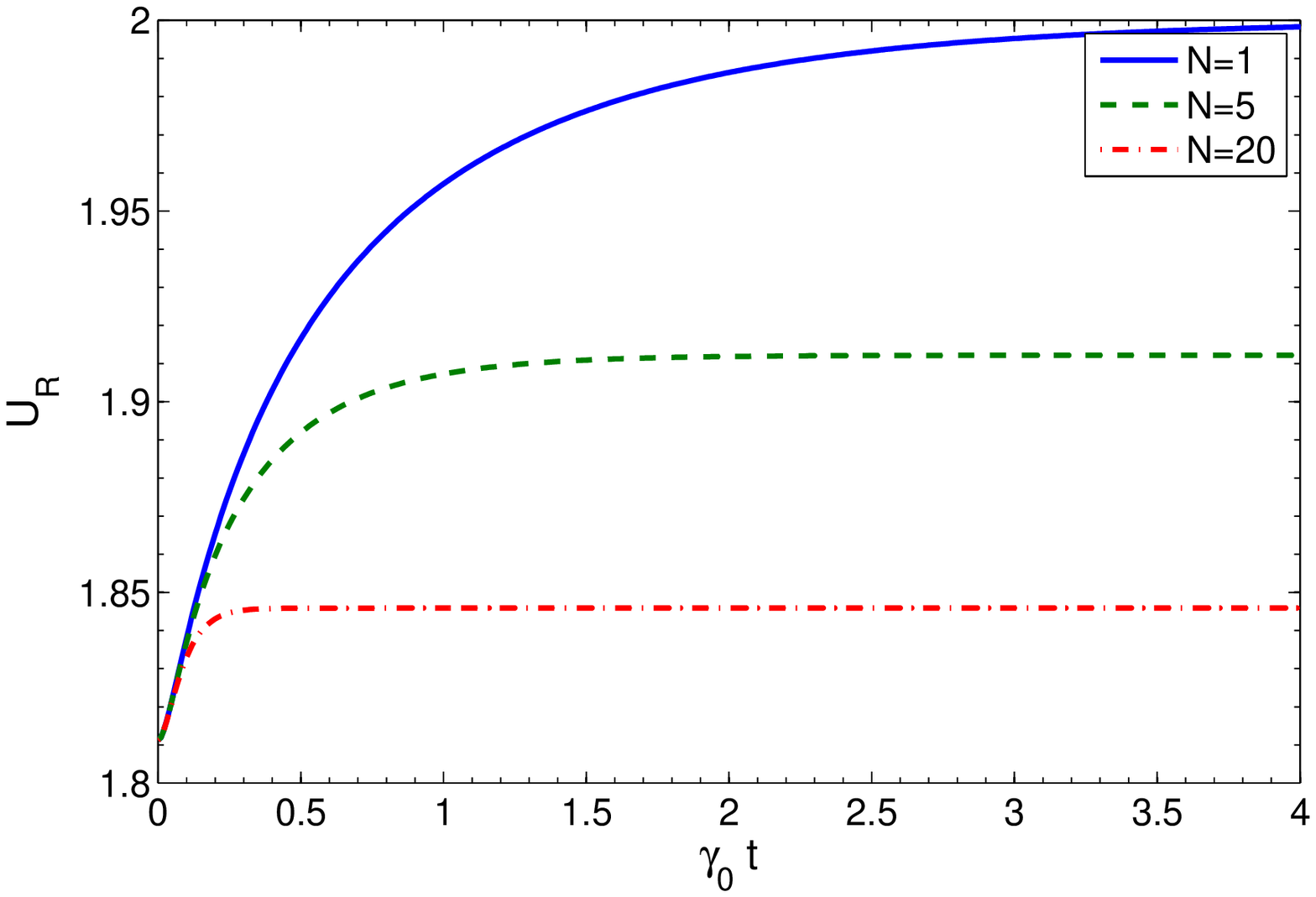}}
\caption{Lower bounds of the entropic uncertainty relation of the two complementary observables $\hat{\sigma}_x$ and $\hat{\sigma}_z$ as a function of $\gamma_0 t$, when Bob prepare a Bell diagonal state s$\rho^{AB_i}=p \vert \psi^- \rangle\langle \psi^- \vert + \frac{1-p}{2}(\vert \psi^+ \rangle\langle \psi^+ \vert + \vert \phi^+ \rangle\langle \phi^+ \vert)$ with $p=1/2$ in Markovian regime $\lambda=40 \gamma_0$. }\label{Fig5}
\end{figure}
Alice and Bob reach an agreement on  measuring two observables $\hat{\sigma_x}$ and $\hat{\sigma_z}$. The von Neumann entropy of the post measurement states are given by
\begin{eqnarray}
S(\rho_{\hat{\sigma_{x}}B_i})&=&- \frac{\vert C_i(t) \vert^{2}}{2}\log_2 \frac{\vert C_i(t) \vert^{2}}{4} \nonumber   \\
&-&\frac{2-\vert C_i(t) \vert^{2}}{2}\log_2 \frac{2-\vert C_i(t) \vert^{2}}{4}, \nonumber \\
S(\rho_{\hat{\sigma_{z}}B_i})&=& - \frac{\vert C_i(t) \vert ^{2}}{8} \log_2 \frac{\vert C_i(t) \vert ^{2}}{8} -\frac{3 \vert C_i(t) \vert ^{2}}{8} \log_2 \frac{3 \vert C_i(t) \vert ^{2}}{8}\nonumber \\
&-&\frac{4 -3 \vert C_i(t) \vert ^{2}}{8}\log_2 \frac{4 -3 \vert C_i(t) \vert ^{2}}{8} \nonumber \\
&-&\frac{4 -\vert C_i(t) \vert ^{2}}{8}\log_2 \frac{4 -\vert C_i(t) \vert ^{2}}{8},
\end{eqnarray}
So, the left-hand side (LHS) of Eq.(\ref{concen}) is obtained as
\begin{eqnarray}
U_L&=&- \frac{\vert C_i(t) \vert^{2}}{2}\log_2 \frac{\vert C_i(t) \vert^{2}}{4} \nonumber   \\
&-&\frac{2-\vert C_i(t) \vert^{2}}{2}\log_2 \frac{2-\vert C_i(t) \vert^{2}}{4} - \frac{\vert C_i(t) \vert ^{2}}{8} \log_2 \frac{\vert C_i(t) \vert ^{2}}{8} \nonumber \\
&-&\frac{3 \vert C_i(t) \vert ^{2}}{8} \log_2 \frac{3 \vert C_i(t) \vert ^{2}}{8}-\frac{4 -\vert C_i(t) \vert ^{2}}{8}\log_2 \frac{4 -\vert C_i(t) \vert ^{2}}{8}\nonumber \\
&-&2 S_{bin}(\frac{\vert C_i(t) \vert ^{2}}{2}),
\end{eqnarray}
The right-hand side (RHS) of Eq.(\ref{concen}) is given by
\begin{eqnarray}
U_R&=&1-(\alpha_- -\theta)\log_2 (\alpha_- -\theta) - (\alpha_+ -\theta)\log_2 (\alpha_+ -\theta) \nonumber \\
&-&(\alpha_- + \theta)\log_2 (\alpha_- + \theta)-(\alpha_+ +\theta)\log_2 (\alpha_+ + \theta) \nonumber \\
&+& \max \lbrace 0 , \delta \rbrace-S_{bin}(\frac{\vert C_i(t) \vert ^{2}}{2}),
\end{eqnarray}
where 
\begin{eqnarray}
\alpha_\pm &=& (2 \pm \vert C_i(t) \vert ^{2} )/2, \nonumber \\
\theta&=&(\sqrt{1-\vert C_I{t}  \vert^{2} + \vert C_i(t) \vert ^{4}})/4,
\end{eqnarray}
 and 
\begin{eqnarray}
\delta&=&(\alpha_- -\theta)\log_2 (\alpha_- -\theta) + (\alpha_+ -\theta)\log_2 (\alpha_+ -\theta) \nonumber \\
&+&(\alpha_- + \theta)\log_2 (\alpha_- + \theta)+(\alpha_+ +\theta)\log_2 (\alpha_+ + \theta) \nonumber \\
&-&S_{bin}(\frac{\vert C_i(t) \vert ^{2}}{2})+S(\rho_{\hat{\sigma_{z}}B_i})+S(\rho_{\hat{\sigma_{x}}B_i}).
\end{eqnarray}
In Fig. \ref{Fig4}, the EULB, $U_R$ is plotted as a function of time in non-Markovian regime $\lambda=0.1 \gamma_0$ for various number of additional qubit $N$. As can be seen from Fig. \ref{Fig4}, due to the interaction between quantum memory and reservoir, EULB is increased through time . One can see, the lower bound decreases by increasing the number of additional qubit . In Fig. \ref{Fig5},  $U_R$ is plotted as a function of time in Markovian regime $\lambda = 40 \gamma_0$ for various number of additional qubits $N$. As can be seen from Fig. \ref{Fig5}, $U_R$ is increased over time, while it is decreased by increasing the number of additional qubits.

\section{Conclusion}\label{Sec4}
In this work we have studied the quantum-memory assisted EULB when quantum memory interacts with reservoir. The model we have considered here is such that the quantum memory along with $N-1$ non-interacting qubits located in common reservoir. We assume that these $N$ qubit independently are coupled to a common reservoir. It is logical to expect that the EULB increases over time. However, we have shown that the EULB can be protected from decoherence by controlling the additional qubits in reservoir. It has been shown that for both Markovian and non-Markovian regime the EULB is decreased by increasing the number of additional qubits.

\end{document}